\documentclass[12pt,preprint]{aastex} 
\newcommand{\degree}{\hbox{$^\circ$}}
\newcommand{\etal}{et\,al.}
\newcommand{\halpha}{H$\alpha$}
\newcommand{\hbeta}{H$\beta$}
\newcommand{\gsim}{\raise0.3ex\hbox{$>$}\kern-0.75em{\lower0.65ex\hbox{$\sim$}}}
\newcommand{\kms}{km\,s$^{-1}$}
\newcommand{\lsim}{\raise0.3ex\hbox{$<$}\kern-0.75em{\lower0.65ex\hbox{$\sim$}}}
\newcommand{\mjpbeam}{\,\,mJy\,Beam$^{-1}$}
\newcommand{\msun}{M$_{\odot}$}
\newcommand{\HI}{H~{\sc i}}
\newcommand{\HII}{H~{\sc ii}}
\newcommand{\snu}{S$_{\nu}$}
\begin{document}     
\slugcomment{Accepted for publication in the Astrophysical Journal}
\title{The Nature of Radio Continuum Emission in the Dwarf Starburst Galaxy 
NGC 625}
\author{John M. Cannon and Evan D. Skillman}
\affil{Department of Astronomy, University of Minnesota,\\ 116 Church St. 
S.E., Minneapolis, MN 55455}
\email{cannon@astro.umn.edu, skillman@astro.umn.edu}

\begin{abstract}

We present new multi-frequency radio continuum imaging of the dwarf starburst 
galaxy NGC\,625 obtained with the Very Large Array.  Data at 20, 6, and 3.6 cm 
reveal global continuum emission dominated by free-free emission, with only 
mild synchrotron components.  Each of the major \HII\ regions is detected; the 
individual spectral indices are thermal for the youngest regions (showing 
largest \halpha\ emission) and nonthermal for the oldest.  We do not detect any
sources that appear to be associated with deeply embedded, dense, young 
clusters, though we have discovered one low-luminosity, obscured source that 
has no luminous optical counterpart and which resides in the region of highest 
optical extinction. Since NGC\,625 is a Wolf-Rayet galaxy with strong recent 
star formation, these radio properties suggest that the youngest star 
formation complexes have not yet evolved to the point where their thermal 
spectra are significantly contaminated by synchrotron emission. The nonthermal 
components are associated with regions of older star formation that have 
smaller ionized gas components.  These results imply a range of ages of the 
\HII\ regions and radio components that agrees with our previous resolved 
stellar population analysis, where an extended burst of star formation has 
pervaded the disk of NGC\,625 over the last $\sim$ 50 Myr. We compare the
nature of radio continuum emission in selected nearby dwarf starburst and 
Wolf-Rayet galaxies, demonstrating that thermal radio continuum emission 
appears to be more common in these systems than in typical \HII\ galaxies with 
less recent star formation and more evolved stellar clusters.

\end{abstract}						

\keywords{galaxies: individual (NGC\,625) --- galaxies: starburst
--- galaxies: dwarf --- radio continuum: galaxies}                  

\section{Introduction}
\label{S1}

Radio continuum emission from star formation regions is an invaluable probe of 
the physical conditions therein.  Radio observations offer us one of the only 
truly extinction-free measures of recent (\lsim\ 30 Myr) star formation. With 
the high sensitivity and angular resolution available with modern 
interferometers [e.g., the Very Large Array (VLA)\footnote{The National Radio 
Astronomy Observatory is a facility of the National Science Foundation operated 
under cooperative agreement by Associated Universities, Inc.}], radio continuum 
observations can be compared with optical and infrared observations at 
comparable physical scales (e.g., from HST), allowing us to discern 
evolutionary correlations between stellar and energetic particle populations.

There are two major components that constitute the radio continuum emission 
from a typical star formation region (see {Condon 1992}\nocite{condon92} for a 
comprehensive overview). First, the thermal free-free emission arises directly
from \HII\ regions. The strength of this component is proportional to the total
number of Lyman continuum photons, and hence to the total number of massive 
stars.  With a radio spectrum characterized by \snu\ $\sim\ \nu^{\alpha}$, pure
optically thin thermal bremsstrahlung emission will show a
nearly flat slope ($\alpha \sim -$0.1).  Second, the nonthermal synchrotron 
component is produced by relativistic electrons accelerated in recent 
supernovae (SNe) explosions and remnants.  This emission will show a more 
negative spectral index ($-$1.2 $\lsim\ \alpha\ \lsim\ -$0.4).  Most 
starbursting galaxies show a radio continuum spectrum dominated by one of 
these components, or some combination thereof \citep{deeg93}; see \S~\ref{S4} 
for further discussion.  

A third and less ubiquitous component of the radio continuum emission from 
star-forming galaxies can be produced by free-free absorption, resulting in a 
positive spectral index ($\alpha >$ 0.0).  Such sources have been identified by
\citet{kobulnicky99c} and \citet{johnson03b} as extremely young, dense, 
heavily-embedded star clusters (or ``ultradense \HII\ regions'').  Since the 
sample of galaxies searched for such regions remains small, the total fraction 
of starbursting systems that contain such extreme star formation regions is not
well known. However, preliminary estimates suggest these sources may enshroud 
$\sim$ 10\% of the O-star populations of starburst galaxies \citep{johnson01}.

If radio continuum emission is observed at multiple frequencies and at matched 
resolutions, we can discern which of the above physical processes dominates
within individual star formation regions. By concentrating on nearby starburst 
galaxies that have resolved stellar and cluster populations in the visual or 
infrared, we can probe the youngest phases of star formation, from 
heavily-embedded star clusters to the supernova remnants resulting from massive
star evolution.  This provides detailed insights into the processes that 
regulate star formation in starburst galaxies, including feedback and cluster 
formation and evolution timescales.  These are important quantities both 
locally and in the high-redshift universe.

Here, we present new VLA observations of the dwarf starburst galaxy NGC\,625.  
This system is a nearby \citep[D $=$ 3.89\,$\pm$\,0.22 Mpc;][]{cannon03} 
prototypical dwarf starburst that has revealed many striking properties in 
recent multiwavelength investigations (see Table~\ref{t1} for a summary of 
basic galaxy parameters). NGC\,625 currently hosts a massive starburst with a 
star formation rate $\sim$ 0.05 \msun\,yr$^{-1}$ and which displays Wolf-Rayet 
(W-R) emission features ({Skillman, C{\^ o}t{\' e}, \& Miller 
2003a,b}\nocite{skillman03b,skillman03a}).  We have performed a recent star 
formation history analysis using spatially resolved stellar HST/WFPC2 
photometry \citep{cannon03}, and find that the current burst is actually 
long-lived (\gsim\ 50 Myr) compared to canonical expectations based on W-R 
star populations (ages \lsim\ 6 Myr; {Conti 1991}\nocite{conti91}; {Schaerer, 
Contini, \& Kunth 1999a}\nocite{schaerer99a}).  This extended burst of star 
formation appears to have disrupted the \HI\ disk, and this system is 
currently undergoing outflow of \HI\ from the major starburst region {(Cannon 
\etal\ 2004a)}\nocite{cannon04a}.  This outflow is also seen in diffuse x-rays
\citep{bomans98} and in \ion{O}{6} absorption from FUSE spectroscopy (Cannon 
\etal\ 2004b, in preparation).  

\placetable{t1}

From the above properties it is clear that the star formation in NGC\,625 is 
violent and is having a dramatic impact on the ISM and potentially on the 
surrounding IGM as well.  Since this galaxy is one of only a small subset of
star-forming dwarfs that demonstrate such extreme properties, it is important
that we explore other observational avenues with which to better understand its
evolution.  Here we use multifrequency VLA radio continuum data to infer the 
dominant emission processes in the major star formation regions, as well as 
various characteristics of the recent star formation in this galaxy.

\section{Observations and Data Reduction}
\label{S2}

We obtained VLA radio continuum imaging using the BnA and CnB arrays on 2003, 
October 9, 10, 11, and 2004, February 7 \& 8 (see Table~\ref{t2}) for program 
AC702.  Due to the low declination of the source ($-$41.4\degree, or a maximum 
elevation at the VLA of $\sim$ 15\degree), only very short transit observations
($\sim$ 3 hours) are possible; while the hybrid arrays produce more balanced
{\it u-v} coverage, the beams are still slightly asymmetrical (typically a 
1.5:1.0 ratio in the declination-right ascension directions).

\placetable{t2}

All reductions were performed using the Astronomical Image Processing System
(AIPS) package. First, interference and bad data were removed.  Flux, gain and 
phase calibrations were then applied, derived from observations of 0542+498 
(3c147; primary calibrator) and 0155-408 (secondary calibrator). The C-band (6 
cm) and X-band (3.6 cm) observations were obtained in two arrays, and these 
{\it u-v} databases were then concatenated. The calibrated {\it u-v} data were 
then imaged and cleaned to produce continuum maps at each frequency that 
balanced resolution and sensitivity. When comparing observations at different 
frequencies to derive spectral indices, it is important that the resulting 
beams are of similar size and position angle. Of course there will be different
coverages in the {\it u-v} plane for each frequency and array, so tapering and 
weighting variations were applied to produce a set of images with nearly 
matched synthesized beams; since the L-band (20 cm) observations have the 
poorest resolution, the C-band and X-band data were tapered to this beam size 
for spectral index work.

The resulting matched-resolution images are discussed further in the next 
sections. The beam sizes are 5.86\arcsec\ $\times$ 3.63\arcsec, 5.31\arcsec\ 
$\times$ 3.63\arcsec, and 4.45\arcsec\ $\times$ 3.63\arcsec\ for the L, C and 
X-band data, respectively.  The corresponding rms noise levels are 
0.036\mjpbeam, 0.016\mjpbeam, and 0.015\mjpbeam.  These rms values provide 
sensitivity to the three major star formation regions in NGC\,625, as well as 
to diffuse radio continuum emission in the surrounding regions. The flux 
densities of each detected radio source were measured via two methods: 
two-dimensional Gaussian fitting was performed in AIPS, and the final images
were used in aperture photometry routines to assure the same physical areas 
were sampled in constructing spectral index measurements.  Source flux 
densities are discussed further in \S~\ref{S3}.

\section{Radio Continuum Emission In NGC\,625}
\label{S3}

\begin{figure}
\epsscale{0.85}
\plotone{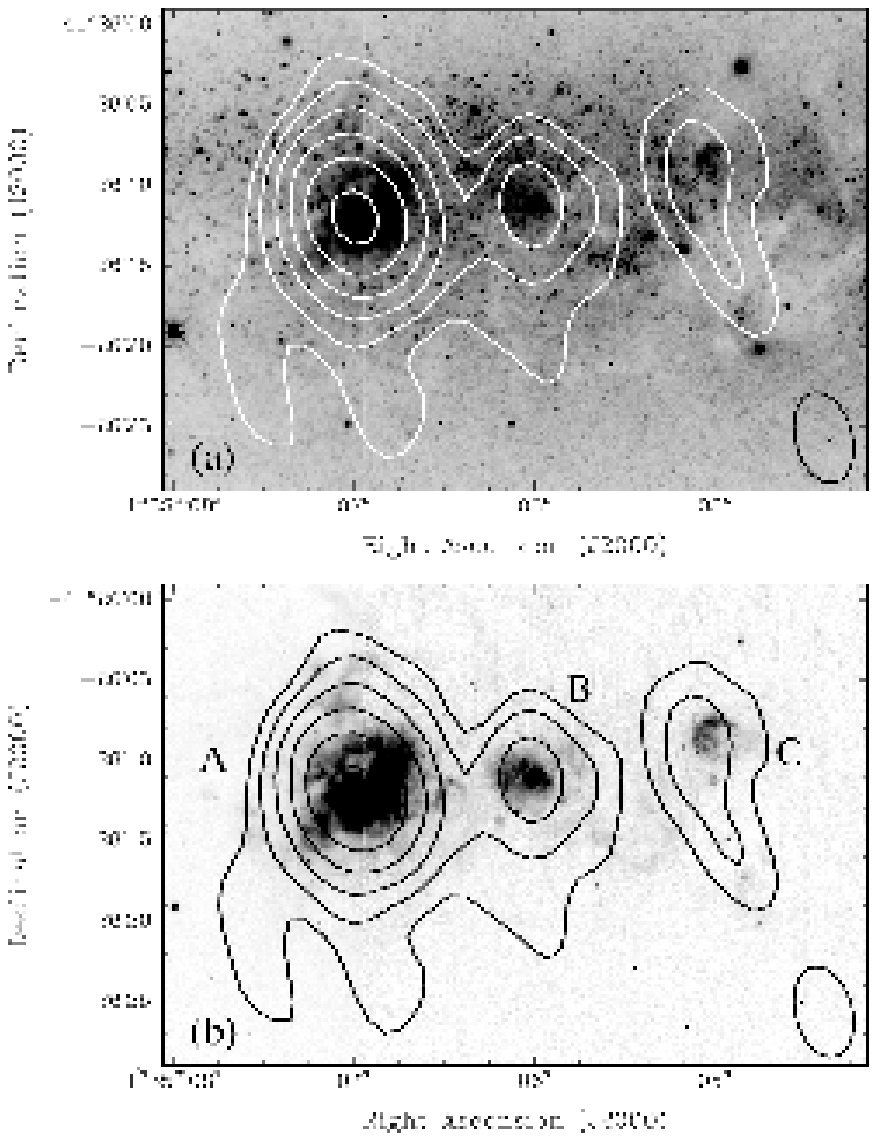}
\caption{L-Band ($\nu$ $=$ 1.40 GHz) emission contours at the 3\,$\sigma$ 
(1.08$\times$10$^{-4}$ Jy\,Beam$^{-1}$), 6\,$\sigma$, 9\,$\sigma$, 
12\,$\sigma$, 15\,$\sigma$, and 18\,$\sigma$ levels, superposed on an HST/WFPC2
F555W (V) image (a) and on an HST/F656N (continuum-subtracted \halpha) image 
(b).  The labels A, B, and C in (b) denote the nomenclature from 
\citet{cannon03}, and mark the three major \HII\ regions.  Note the excellent 
spatial agreement between \halpha\ emission and radio continuum peaks. 
The beam size is 5.86\arcsec\ $\times$ 3.63\arcsec, and is shown at bottom
right.}
\label{figcap1}
\end{figure}

\begin{figure}
\epsscale{0.85}
\plotone{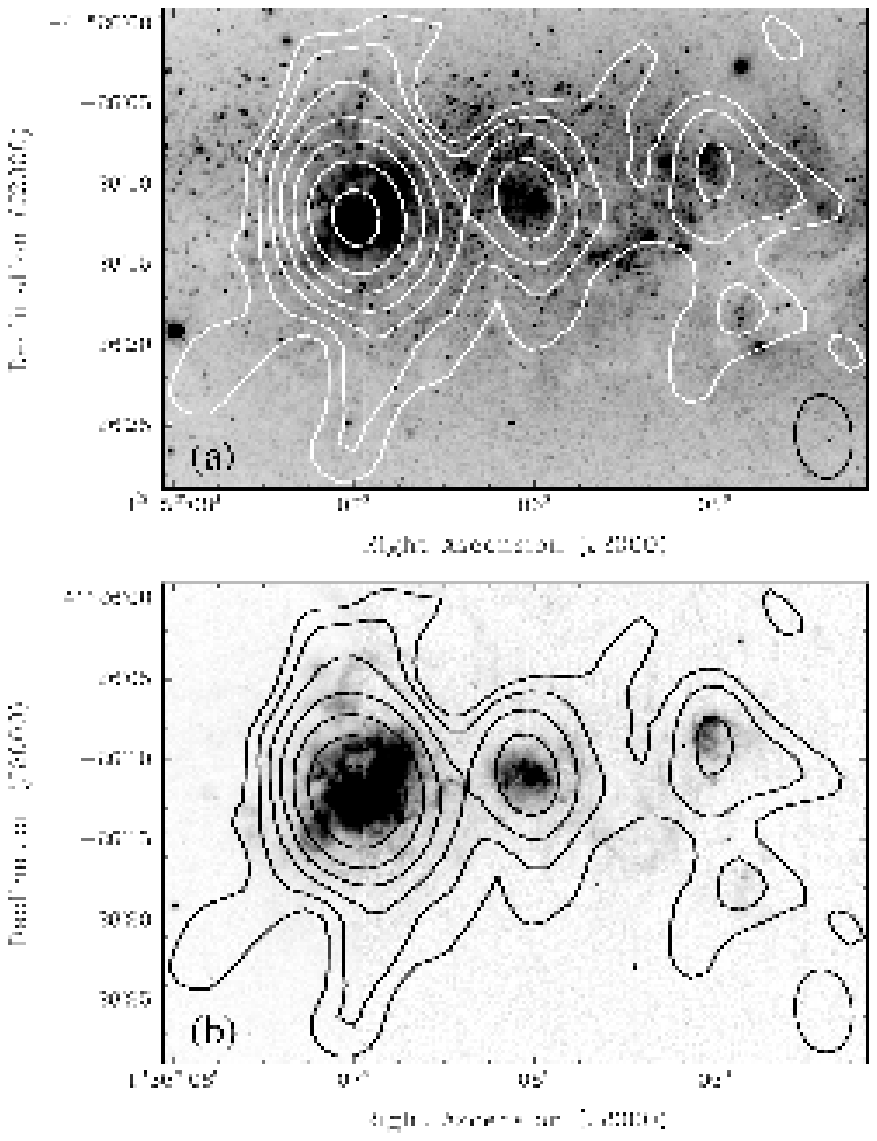}
\caption{C-Band ($\nu$ $=$ 4.86 GHz) emission contours at the 3\,$\sigma$ 
(4.8$\times$10$^{-5}$ Jy\,Beam$^{-1}$), 6\,$\sigma$, 9\,$\sigma$, 
12\,$\sigma$, 15\,$\sigma$, 18\,$\sigma$ and 21\,$\sigma$ levels, superposed 
on an HST/WFPC2 F555W (V) image (a) and on an HST/F656N (continuum-subtracted 
\halpha) image (b).  The beam size is 5.31\arcsec\ $\times$ 3.63\arcsec, and 
is shown at bottom right.}
\label{figcap2}
\end{figure}

\begin{figure}
\epsscale{0.85}
\plotone{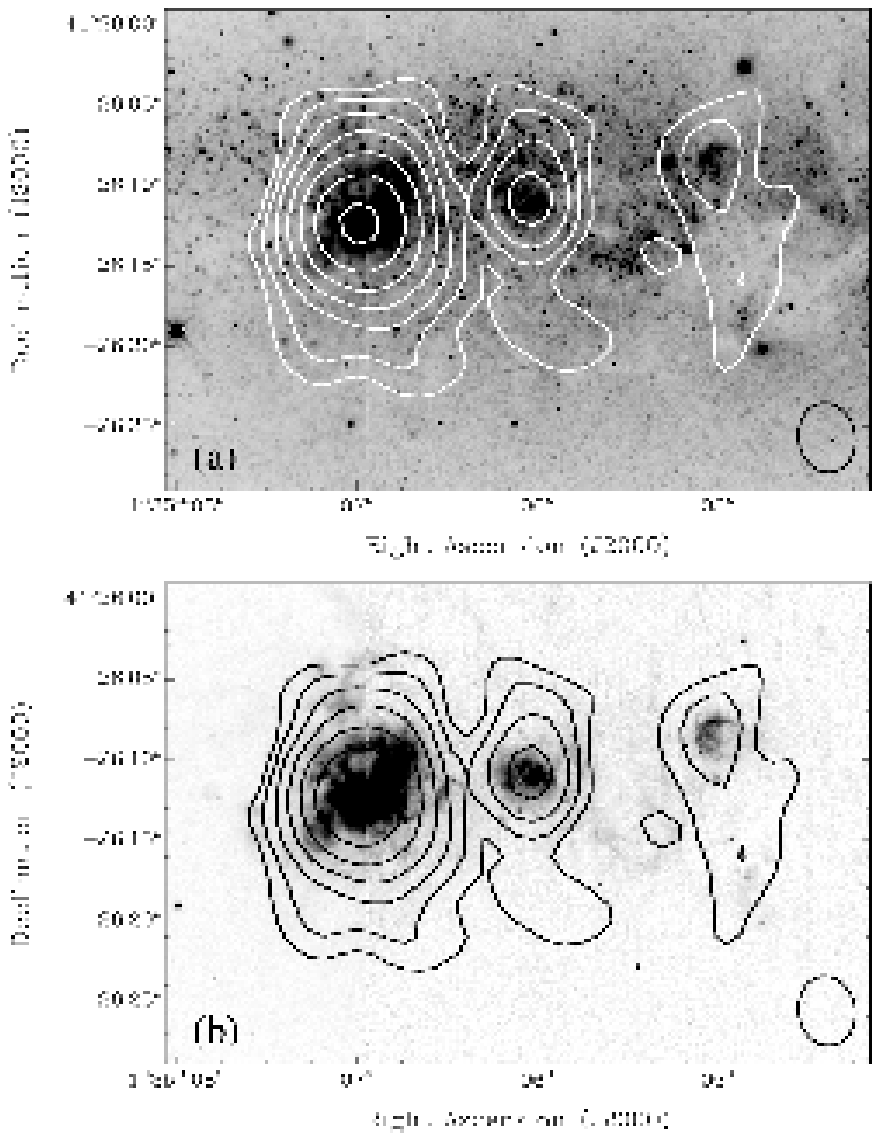}
\caption{X-Band ($\nu$ $=$ 8.46 GHz) emission contours at the 3\,$\sigma$ 
(4.5$\times$10$^{-5}$ Jy\,Beam$^{-1}$), 6\,$\sigma$, 9\,$\sigma$, 
12\,$\sigma$, 15\,$\sigma$, 18\,$\sigma$, and 21\,$\sigma$ levels, superposed 
on an HST/WFPC2 F555W (V) image (a) and on an HST/F656N (continuum-subtracted 
\halpha) image (b). The beam size is 4.45\arcsec\ $\times$ 3.63\arcsec, and is 
shown at bottom right.}
\label{figcap3}
\end{figure}

We detect radio continuum emission from the three largest star formation 
regions (as traced by \halpha\ luminosity) in NGC\,625.  In 
Figures~\ref{figcap1}, \ref{figcap2}, and \ref{figcap3} we present overlays of 
the L, C, and X-band images on HST/WFPC2 V and (continuum-subtracted) \halpha\ 
images (see {Cannon \etal\ 2003}\nocite{cannon03} for a detailed analysis of 
these images).  There is an excellent correlation between \halpha\ emission and
radio continuum emission at this sensitivity level [note, however, that 
{Skillman \etal\ (2003a)}\nocite{skillman03a} find diffuse \halpha\ emission 
throughout the disk and extending above the plane to the north].  There is 
also a good correspondence between massive stellar clusters and radio continuum
peaks (see below).  

In a previous analysis of the WFPC2 images used for comparison here, we found a
position offset of 5.1\arcsec\ in declination only from a ground-based 
astrometric plate solution \citep[see][]{cannon03}.  The source of this error 
is unknown, and we have corrected the positions in this work.  The positional 
alignment accuracy between the radio and the HST optical images is dominated by
the uncertainty in the HST coordinates ($\sim$ 0.1\arcsec\ rms in the radio, 
and $\sim$ 0.5\arcsec\ rms for HST).  We conservatively estimate that the 
positional accuracy between the two datasets is 0.6\arcsec\ rms.  At the 
distance of 3.89 Mpc, this corresponds to 11 pc, or greater than the radii of 
many young, dense stellar clusters.  This will preclude us from assuring 
accurate comparisons between the smallest stellar clusters and radio continuum 
emission peaks; however, as described below, the close morphological agreement 
between radio and \halpha\ emission suggests our astrometric solution is 
robust.  

To measure total flux densities in each of the observing bands, all emission 
detected above the 3\,$\sigma$ level surrounding the optical region of the 
galaxy was integrated within an identical aperture for each frequency.  These 
values are given in Table~\ref{t3}. The global spectral index of NGC\,625 is 
nearly purely thermal, with $\alpha = -$0.13\,$\pm$\,0.08. This suggests that 
the bulk of the radio continuum emission arises from current \HII\ regions, 
with the number of Lyman continuum photons proportional to the total number of 
massive stars (see below).  

Since care was taken to achieve matching beam sizes at all three observing 
frequencies, we can also compare the spectral indices of individual \HII\ 
regions and radio continuum peaks.  As discussed in \S~\ref{S2}, the flux 
densities of individual peaks were measured both by two-dimensional Gaussian 
fitting and by aperture photometry with fixed aperture size. As an inspection 
of Figures~\ref{figcap1}, \ref{figcap2} and \ref{figcap3} will reveal, there 
exists diffuse radio continuum emission throughout the disk at this resolution.
This will complicate the derivation of flux densities from individual peaks, 
since each frequency attains a different noise level and hence surface 
brightness sensitivity.  For this reason we derive source flux densities using 
aperture photometry with matched aperture sizes between images.  This assures 
that only emission associated directly with each emission peak is included for 
the derivation of spectral indices.  The results for each frequency are 
presented in Table~\ref{t3}.

\subsection{Thermal Radio Continuum Sources}
\label{S3.1}

The two highest-surface brightness \HII\ regions, NGC\,625\,A and NGC\,625\,B 
(labeled in Figure~\ref{figcap1}b), are found to have thermal spectral indices 
with $\alpha \simeq -$0.10. These radio maxima are thus powered by free-free 
emission.  Since these peaks have unambiguously flat spectral indices, we use 
their thermal emission to estimate the number of ionizing stars in the 
associated \HII\ regions, as well as the current star formation rates.  From 
\citet{condon92}, for a purely thermal source, the number of Lyman continuum 
photons is approximated by:

\begin{equation}
N_{UV} \gsim\ 6.3\times10^{52} (T_e)^{-0.45} (\nu)^{0.1} L_T
\end{equation}

\noindent where N$_{UV}$ is the number of Lyman continuum photons emitted per 
second, T$_e$ is the electron temperature in units of 10$^4$ K, $\nu$ is the 
observing frequency in GHz, and L$_T$ is the thermal radio continuum 
luminosity in units of 10$^{20}$ W\,Hz$^{-1}$.  Also from \citet{condon92}, the
thermal radio continuum flux density is related to the current star formation 
rate by the relation:

\begin{equation}
SFR \sim\ \frac{L_T \cdot\ \nu^{0.1}}{5.5\times10^{20}}
\end{equation}

\noindent where SFR is the current star formation rate of stars more massive 
than 5\,\msun\ in units of \msun\,yr$^{-1}$, L$_T$ is the thermal radio 
continuum luminosity in units of W\,Hz$^{-1}$, and $\nu$ is the observing 
frequency in GHz.

Using equation (1) above and adopting the electron temperatures for regions 
NGC\,625\,A and NGC\,625\,B from {Skillman \etal\ (2003b}\nocite{skillman03b}; 
their \HII\ regions 5, 9 and 18 correspond to NGC\,625 A, B, C, 
respectively; see also Table~\ref{t3} and \S~\ref{S3.4}) and the distance of 
3.89 Mpc from 
\citet{cannon03}, we find that \HII\ regions NGC\,625\,A and NGC\,625\,B 
together emit $\sim$ 10$^{52}$ Lyman continuum photons per second.  Using 
conversions from \citet{vacca94} and {Vacca, Garmany, \& Shull 
(1996)}\nocite{vacca96}, where one O7\,V star emits 10$^{49}$ Lyman continuum 
photons per second, this implies that these two \HII\ regions harbor $\sim$ 
1000 such O stars (see Table~\ref{t3}).  

Similarly, using equation (2) we estimate the current massive star formation 
rate in these two \HII\ regions to be $\sim$ 0.03 \msun\,yr$^{-1}$ (see 
Table~\ref{t3}). Since these two \HII\ regions constitute $\sim$ 98\% of the 
\halpha\ flux associated with high-surface brightness \HII\ regions in NGC\,625
and $\sim$ 95\% of the total \halpha\ from the galaxy \citep[see discussion 
in][]{cannon03}, this estimate of the star formation rate encompasses nearly 
all of the area undergoing active star formation at the current epoch.  It is 
in agreement with, though slightly less than, the estimate of 0.05 
\msun\,yr$^{-1}$ found from \halpha\ fluxes in \citet{skillman03a} and 
\citet{cannon03}; the difference is due to assumptions about the initial mass 
function used in each method (the radio method giving the star formation rate 
for stars more massive than 5\,\msun, and the \halpha\ method integrated over 
a 0.1 - 100 \msun\ range). 

\subsection{Nonthermal Radio Continuum Sources}
\label{S3.2}

\HII\ region NGC\,625\,C is found to have a more negative spectral index, 
$\alpha = -$0.33\,$\pm$\,0.08.  This suggests that nonthermal processes are 
more dominant in this region compared to regions NGC\,625\,A and NGC\,625\,B
discussed above.  Synchrotron radiation from relativistic electrons likely 
arises in SNe explosions and remnants.  Thus the predominance of this type of 
emission suggests that these regions are somewhat older than thermal regions
(see below), and that the cumulative contribution from the evolution of massive
stars is more pronounced. 

NGC\,625\,C is the least-luminous \HII\ region detected in our HST/WFPC2 
\halpha\ imaging. From an examination of the \halpha\ morphology in 
Figures~\ref{figcap1}(b), \ref{figcap2}(b), and \ref{figcap3}(b), the 
high-surface brightness areas of NGC\,625\,C have a shell morphology, with the 
eastern side more luminous than the western.  Such a scenario likely results 
when numerous SNe detonate in a localized area in a short time interval.  The 
size of the putative shell ($\sim$ 50 pc) greatly exceeds the sizes of typical 
single-star SN remnants and implies that numerous massive stars had to once
power this star formation region in order to create such a large shell of 
ionized gas.  The nonthermal radio continuum emission arising from SNe remnants
associated with this star formation region implies that it is older than the 
thermal regions. 

These results are consistent with our HST/WFPC2 star formation history analysis
\citep[see][]{cannon03}, where widespread star formation was found throughout 
the disk over at least the last $\sim$ 50 Myr.  In that investigation, local 
star formation concentrations (as probed by the blue helium burning stars, 
which are useful, luminous age probes) were shown to propagate throughout the 
disk over the last $\sim$ 100 Myr, with a general movement from west to east
[see in particular Figure~14 of {Cannon \etal\ (2003)}\nocite{cannon03}]. 
Since radio continuum emission is only associated with massive stars with 
lifetimes \lsim\ 30 Myr \citep{condon92}, these observations do not allow us 
to probe this entire epoch. However, the mix of thermal and nonthermal 
components is consistent with star formation over the last 30 Myr, in various 
areas of the disk, in agreement with our HST results.  Furthermore, we find 
that, in general, the radio emission peaks are more nonthermal in the western 
park of the disk, and more thermal in the eastern part.  As argued in 
\S~\ref{S4}, this is consistent with an age progression from older to younger,
in general agreement with the star formation history analysis. 

\subsection{Deeply Embedded Star Formation Regions?}
\label{S3.3}

\begin{figure}
\epsscale{0.85}
\plotone{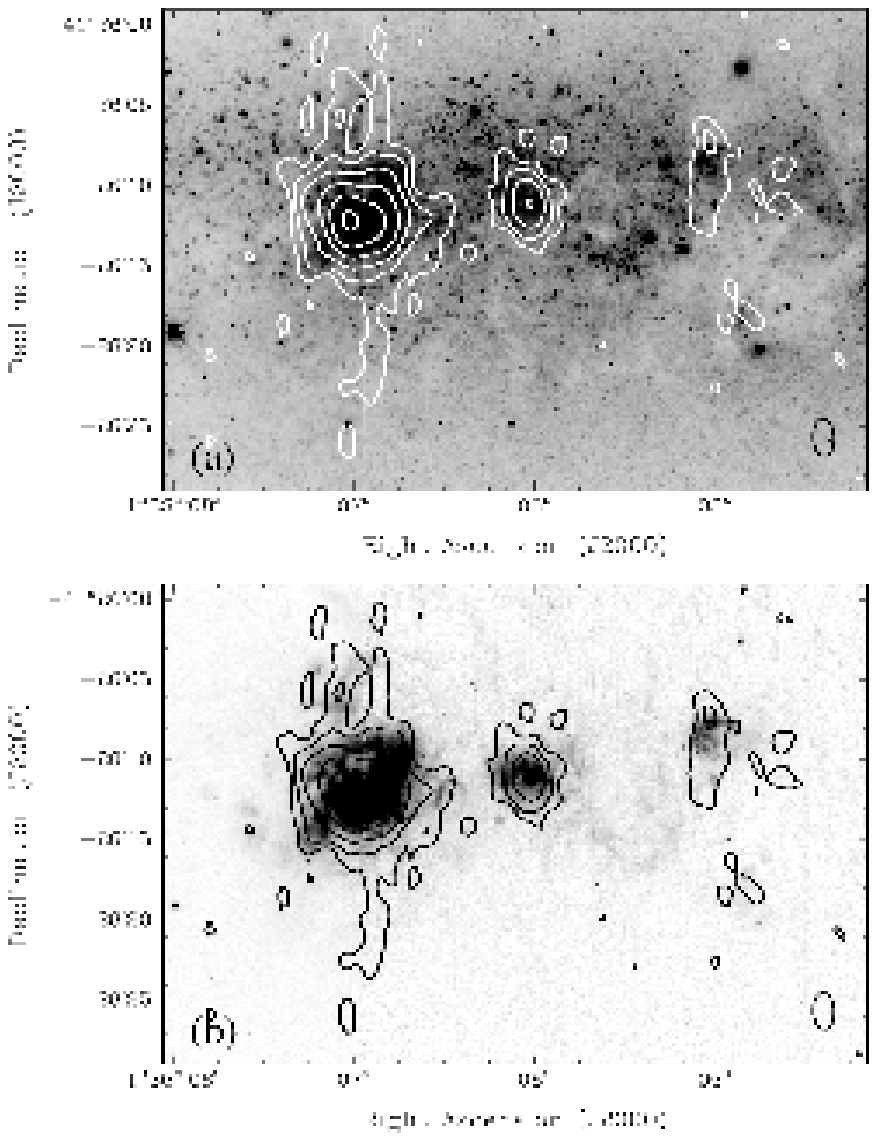}
\caption{Higher-resolution C-Band ($\nu$ $=$ 4.86 GHz) emission contours at the
3\,$\sigma$ (5.1$\times$10$^{-5}$ Jy\,Beam$^{-1}$), 6\,$\sigma$, 9\,$\sigma$, 
12\,$\sigma$, 15\,$\sigma$, 18\,$\sigma$ and 21\,$\sigma$ levels, superposed 
on an HST/WFPC2 F555W (V) image (a) and on an HST/F656N (continuum-subtracted 
\halpha) image (b).  At this resolution we do not detect any deeply-embedded 
sources with positive spectral index (see further discussion in \S~\ref{S3.3}.)
The beam size is 2.4\arcsec\ $\times$ 1.4\arcsec, and is shown at bottom
right.}
\label{figcap4}
\end{figure}

\begin{figure}
\epsscale{0.85}
\plotone{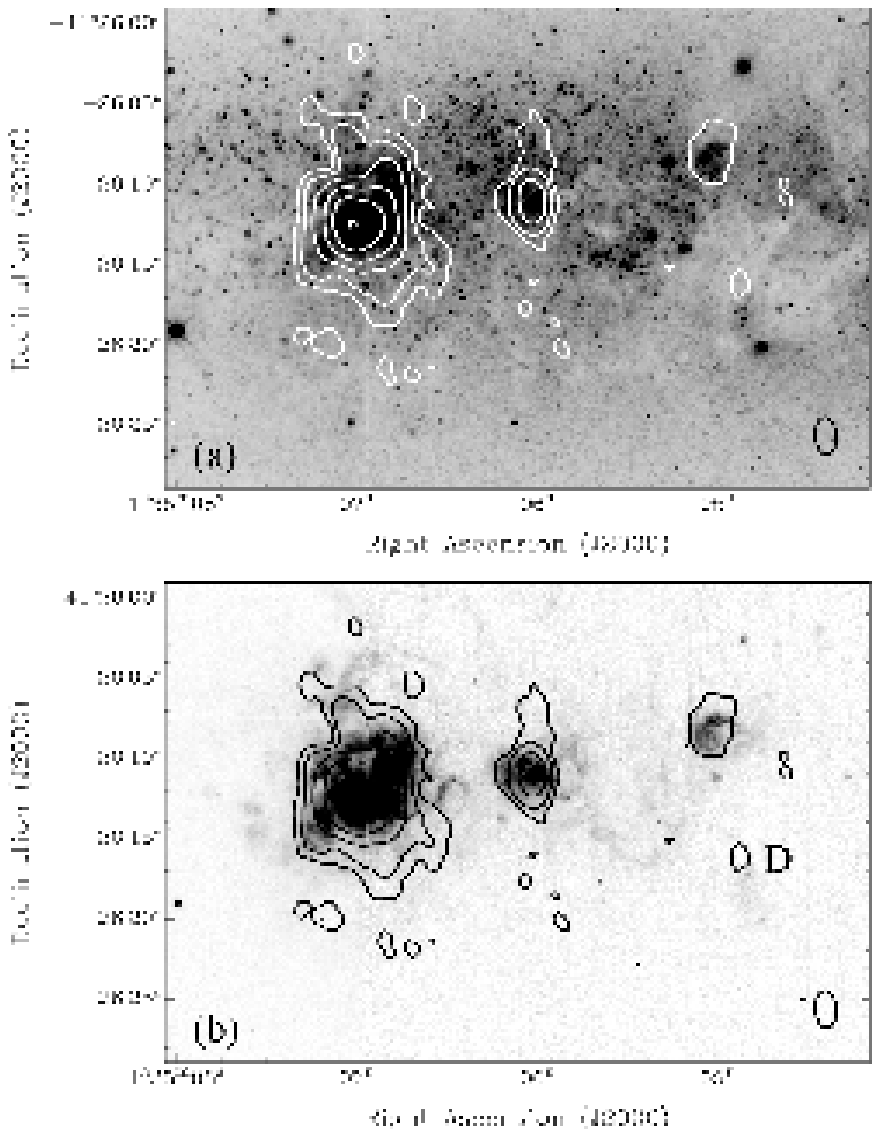}
\caption{Higher-resolution X-Band ($\nu$ $=$ 8.46 GHz) emission contours at the
3\,$\sigma$ (4.8$\times$10$^{-5}$ Jy\,Beam$^{-1}$), 6\,$\sigma$, 9\,$\sigma$, 
12\,$\sigma$, 15\,$\sigma$, 18\,$\sigma$, and 21\,$\sigma$ levels, superposed 
on an HST/WFPC2 F555W (V) image (a) and on an HST/F656N (continuum-subtracted 
\halpha) image (b).  At this resolution we do not detect any deeply-embedded 
sources with positive spectral index (see further discussion in 
\S~\ref{S3.4}), although source ``D'' [labeled in part (b)] is just resolved 
in C and X-bands; this source shows a moderately nonthermal spectral index of 
$-$0.22\,$\pm$\,0.19. The beam size is 2.4\arcsec\ $\times$ 1.4\arcsec, and is 
shown at bottom right.}
\label{figcap5}
\end{figure}

With the high angular resolution afforded by the BnA array at C and X-bands, we
can perform a search for deeply-embedded and obscured star formation regions 
in NGC\,625. A second set of maps was produced at each of these frequencies, 
with matched resolutions of 2.4\arcsec\ $\times$ 1.4\arcsec\ and rms noise 
levels of 0.017\mjpbeam\ and 0.016\mjpbeam\ at C and X-bands, respectively. 
These images, shown in Figures~\ref{figcap4} and \ref{figcap5}, were searched 
for compact sources that may have been lost in the lower-resolution maps.

We find no sources showing positive spectral index values, which are indicative
of embedded stellar clusters still in their natal material \citep{johnson01}.  
However, we do detect one partially optically obscured, slightly nonthermal 
source located within the large dust cloud to the south and west of the three 
radio continuum peaks seen in the lower-resolution maps.  This source, labeled
D in Figure~\ref{figcap5}(b), was then isolated in the low-resolution images 
and its flux density measured in all three observing bands.  This procedure 
yields a spectral index of $-$0.22\,$\pm$\,0.19, indicative of a mix of thermal 
and nonthermal components, and of an older age for its associated cluster.  Due 
to its low luminosity (and thus large error on the implied spectral index), we 
do not over-interpret this detection.  However, taken at face value, it supports
the idea of widespread star formation throughout the disk over the last few 
tens of millions of years, and suggests that the major dust concentration may 
also harbor plentiful molecular gas.  Indeed, this position is coincident with
a CO (2$\rightarrow$1) detection, and observations are planned to map the 
entire system in CO lines (C{\^ o}t{\' e}, Braine, \& Cannon, in preparation).

These higher-resolution maps probe structures smaller than $\sim$ 45 pc at the
distance of NGC\,625. This is the highest matched-beam resolution available at 
these frequencies for this galaxy, due to its southern declination.  A search 
at higher frequencies could be fruitful, with the advantage in resolution; 
however, such observations will be difficult given the elevation-dependent 
effects inherent at those observing bands\footnote{See the Very Large Array 
Observational Status Summary at http://www.vla.nrao.edu/astro/}.  Comparing the
flux densities of sources NGC\,625\,A, NGC\,625\,B, and NGC\,625\,C as measured
with three frequencies at lower resolution (discussed above and presented in 
Table~\ref{t3}) and these higher-resolution data (with only two frequencies),
we find the differences to be smaller than the uncertainties.  Since the formal
errors are smaller on the lower-resolution data (due to the larger wavelength
baseline), we quote these values as characteristic of the more luminous \HII\ 
regions.  

\subsection{Extinction Toward The \HII\ Regions}
\label{S3.4}

By comparing thermal radio continuum flux densities with recombination line 
intensities (e.g., \halpha, \hbeta), one can gauge the reddening toward sources 
that may be partially or completely obscured at optical wavelengths. The wide 
wavelength separation between the optical and radio allows sources that appear 
to have little optical extinction (e.g., by comparing relatively closely-spaced 
optical recombination lines) to be tested for reddening within the system. This 
method fails for strongly nonthermal sources, so we constrain our discussion to 
those \HII\ regions where we have calculated mostly flat radio continuum 
spectral indices, i.e., regions NGC\,625\,A, NGC\,625\,B, and potentially 
NGC\,625\,D (see below).

We compare flux densities derived from both the low and the high-resolution 
X-band observations (the highest-sensitivity frequency presented here) to our 
previously-published HST/WFPC2 \halpha\ fluxes \citep[see][]{cannon03}. From
\citet{caplan86}, for a purely thermal radio continuum source in the absence of
extinction, the \halpha\ flux and the flux density at 8.46 GHz are related via:

\begin{equation}
\frac{j_{H\alpha}}{j_{8.46}} = 
\frac{(8.66\times10^{-9})(T)^{-0.44}}{(8.68 + 
1.5\cdot ln(T))}
\end{equation}

\noindent where j$_{H\alpha}$ is the flux at \halpha\ in 
erg\,sec$^{-1}$\,cm$^{-2}$, j$_{8.46}$ is the flux density at 8.46 GHz in Jy, 
and T is the electron temperature in units of 10$^4$ K. We adopt the 
temperatures for the \HII\ regions from \citet[][see 
Table~\ref{t3}]{skillman03b}. For source
NGC\,625\,D, no temperature is available, so we assume 10$^4$ K, typical of 
the other \HII\ regions. Given these temperatures, we expect 
\begin{math}\frac{j_{H\alpha}}{j_{1384}}\end{math} $=$ 
9.5$\times$10$^{-10}$ erg sec$^{-1}$ cm$^{-2}$ Jy$^{-1}$ for NGC\,625\,A, 
\begin{math}\frac{j_{H\alpha}}{j_{1384}}\end{math} $=$ 
9.7$\times$10$^{-10}$ erg sec$^{-1}$ cm$^{-2}$ Jy$^{-1}$ for NGC\,625\,B, 
and \begin{math}\frac{j_{H\alpha}}{j_{1384}}\end{math} $=$ 
1.0$\times$10$^{-9}$ erg sec$^{-1}$ cm$^{-2}$ Jy$^{-1}$ for NGC\,625\,D.  
As shown in Table~\ref{t3}, our calculated value for 
\begin{math}\frac{j_{H\alpha}}{j_{1384}}\end{math} (which is the average of the
low- and high-resolution values, which agree to within 10\%) for NGC\,625\,B is
consistent with no additional reddening toward this \HII\ region.  NGC\,625\,A,
on the other hand, suffers from mild extinction at \halpha, A$_{H\alpha} =$ 
1.1\,$\pm$\,0.3 magnitudes. The radio continuum source NGC\,625\,D is only 
weakly detected in \halpha\ in the HST imaging; if its radio emission is 
thermal (allowed within the errors), then the impied extinction is a lower
limit of $>$ 2.0 magnitudes.  This source appears to be coincident with \HII\
region \#17 from \citet{skillman03a}; the \halpha\ flux found therein is larger
than our value by $\sim$ 50\%.  If this latter value correct (implying that 
the lower surface brightness sensitivity of the HST narrowband filter has 
missed a substantial fraction of this emission), then the implied extinction
falls to \gsim\ 1.1 magnitudes.

These implied extinction values for regions NGC\,625\,A and NGC\,625\,B are 
fully consistent with those found from 1384 MHz radio continuum imaging 
presented in {Cannon \etal\ (2004a)}\nocite{cannon04a}.  These results enforce
the conclusions drawn from our HST/WFPC2 star formation history analysis, 
where pronounced and variable extinction was found throughout the disk of this
system.  Deep infrared imaging would be insightful to probe the de-reddened 
stellar population behind the large dust concentration.

\subsection{Diffuse Continuum Emission}
\label{S3.5}

We can estimate the flux density of diffuse radio continuum emission as the 
difference between the global radio flux density and the flux densities 
enclosed within the discrete regions discussed above.  However, it should be 
noted that these interferometric observations are biased toward compact,
high-surface brightness regions and that a large fraction of putative diffuse 
emission could be resolved out with these data alone.  With these cautions in 
mind, we estimate that the spectral index of the diffuse emission is 
nonthermal, $\alpha = -$0.49\,$\pm$\,0.69.  Note that the errors on this index 
are large and are dominated by the errors in the measures of the individual 
peak flux densities. 

From our previous Australia Telescope Compact Array (ATCA)\footnote{The 
Australia Telescope is funded by the Commonwealth of Australia for operation as
a National Facility managed by the Commonwealth Scientific and Industrial 
Research Organisation} radio continuum imaging at 1384 MHz (see Cannon \etal\ 
2004a)\nocite{cannon04a}, we find a total L-band flux density of 
10.3\,$\pm$\,1.0 mJy, i.e., within 5\% of the measured global flux density from
the present observations.  Since the ATCA data had shorter baselines than the 
shortest included here (by roughly a factor of two), the close agreement 
implies that the diffuse emission may account for up to $\sim$ 10\% of the 
measured flux density in each band.  This diffuse emission may be the result of
the extended star formation that has been ongoing for the last $\sim$ 50 Myr in
this system. We suggest that the strong thermal emission peaks are superposed 
on a diffuse nonthermal background; the higher thermal luminosities make it 
more difficult to accurately gauge the total nonthermal budget (see further 
discussion in \S~\ref{S4.1}). 

\section{Comparison With Nearby Star-Forming Galaxies}
\label{S4}

As argued above, we interpret the thermal and nonthermal components of the 
radio continuum emission in NGC\,625 as an age progression. As a general rule, 
the slope of the radio spectral index (where \snu\ $\sim$ $\nu^{\alpha}$) 
appears to be useful as a rough guide to the timescale for recent star 
formation.  Heavily-embedded clusters with ages \lsim\ 1 Myr will appear as 
sources with positive values of $\alpha$, suggesting optically thick thermal 
bremsstrahlung emission. As the slope flattens, thermal free-free emission will
dominate, and the radio continuum peaks will show a good correlation with 
\halpha\ emission. Since the thermal radio flux density is proportional to the 
total number of Lyman continuum photons, this emission will have a similar 
timescale to most \HII\ regions ($\sim$ 10 Myr). Finally, negative spectral 
index values indicate a nonthermal electron population where the cumulative 
effects of SNe explosions and remnants are most pronounced.  These effects 
apparently become dominant on timescales longer than \HII\ region lifetimes, 
although a nonthermal population will be introduced after the evolution of the 
most massive star in a given cluster \citep[e.g.,][]{israel88b}. In the next 
sections we compare the nature of radio continuum emission in dwarf starburst 
and Wolf-Rayet galaxies in the literature, with these timescale arguments in 
mind. 

\subsection{Radio Continuum Emission in Nearby Dwarf Starburst Galaxies}
\label{S4.1}

The small number of nearby dwarf starburst systems limits our understanding of
the radio continuum properties of this important subclass of star-forming 
galaxies.  Since these systems are efficient avenues of massive star formation 
that can have dramatic impacts on the surrounding ISM and IGM, it is important 
that the properties be compared among those systems that have been studied at 
high resolution and sensitivity.  Here, we present a brief comparison of the 
continuum properties of NGC\,625 with those of four other well-studied dwarf 
starbursts, NGC\,1569, NGC\,1705, NGC\,5253, and He\,2-10.  This small but 
growing sample of dwarf starbursts (which is not meant to be comprehensive nor
exclusive) is providing detailed insights into the nature of star formation in 
these systems.  

As discussed above, the radio continuum emission in NGC\,625 is dominated by 
regions showing thermal emission, although the diffuse emission appears to be 
nonthermal in nature. Individual star formation regions display a mix of 
thermal and nonthermal emission, which we have interpreted as an age sequence
for the three highest-surface brightness \HII\ regions.  This is consistent 
with our resolved stellar populations study \citep{cannon03} and suggests that 
active star formation has been present throughout the disk for at least the 
last 30 Myr (see discussion in \S~\ref{S3.2}).

At a distance of 2.2 Mpc \citep{israel88a}, NGC\,1569 is the closest known 
dwarf starburst galaxy.  \citet{israel88b} find a nonthermal halo superposed
on a disk dominated by thermal emission, much as seen in the present 
observations.  Those authors also infer NGC\,1569 to be a ``post-starburst''
system, which has been borne out by subsequent HST studies 
\citep[e.g.,][]{greggio98}.  Higher-resolution observations presented in 
\citet{greve02} and \citet{lisenfeld04} find a largely thermal radio 
continuum, with the ``super star clusters'' (SSCs) being mostly thermal in 
nature.  The thermal radio emission peaks coincide with the \halpha\ maxima 
and not the stellar clusters, showing the locations of the most recent star 
formation.  This offset is interpreted as the effect of feedback, where the 
recent star formation that produced the SSCs has cleared out the regions 
surrounding them. 

NGC\,5253 (D $=$ 4.1\,$\pm$\,0.2; {Saha \etal\ 1995}\nocite{saha95}) also 
displays strong thermal radio continuum emission, though again the peaks are 
not directly coincident with the major stellar clusters ({Turner, Ho, \& Beck 
1998}\nocite{turner98}). This high thermal fraction is interpreted as evidence 
for the youth of the current starburst.  Diffuse nonthermal emission surrounds 
these central thermal peaks, and there also exist inverted-spectrum ($\alpha >$
0.0) sources that higher-frequency observations revealed to be dense, compact 
($\sim$ few pc) \HII\ regions ({Turner, Beck, \& Ho 2000}\nocite{turner00}). 
These features suggest that NGC\,5253 is currently forming massive stars at a 
prodigious rate, and that pressures and densities are conducive to the 
formation of potential SSCs that are still deeply-embedded in their birth 
material. 

At a larger distance of 9 Mpc, He\,2-10 is found to demonstrate a strongly 
nonthermal global spectral index ($\alpha$ $\simeq$ $-$0.5; {Kobulnicky \&
Johnson 1999}\nocite{kobulnicky99c}). However, each of the main radio continuum
emission peaks shows either flat or inverted spectral indices.  This is 
interpreted as the signature of very young (age $<$ 1 Myr) stellar clusters 
still in the process of formation \citep[see also][]{johnson03b}. The fact that
He\,2-10 contains such a high fraction of young, embedded potential SSCs 
suggests that the conditions are optimal for the formation of these relatively 
rare, massive young stellar clusters.  

From the above small sample, it appears that radio continuum emission is 
common in strongly star-forming dwarf galaxies, especially those that have 
undergone recent massive cluster formation (i.e., SSCs or their precursors).  
However, NGC\,1705, well-known for its intense galactic wind and prominent 
outflow \citep{heckman01a}, is not detected in the radio continuum at a 
sensitivity level sufficient to detect a typical Galactic ultradense \HII\ 
region {(Johnson, Indebetouw, \& Pisano 2003)}\nocite{johnson03a}.  This is 
interpreted as a recent and rapid termination of star formation in this system,
potentially by the current outflow as it carries energy and perhaps parts of 
the ISM away from the starburst regions.  Based on the lifetime of the massive 
stellar clusters ($\sim$ 1--8 Myr; {Tremonti \etal\ 2001}\nocite{tremonti01}), 
this termination must have been abrupt indeed.

The above sample suggests that most, but not all, systems with vigorous current
star formation (i.e., within the last few Myr) will show strong radio continuum
emission.  On the other hand, systems that have efficiently formed stars over 
the recent epoch may quickly become undetectable in the radio continuum.  The 
timescales and mechanisms for the decrease in radio continuum flux density are 
not well-constrained by observations, primarily because the evolution of radio 
continuum emission processes is intricately linked to the process of feedback 
in dwarf galaxies.  This is an especially difficult quantity to place useful 
observational constraints on, since there are numerous complicated effects 
(e.g., turbulence) that have yet to be understood.  We present in Table~\ref{t4}
a brief compilation of important properties that may affect the interpretation
of radio continuum emission in these systems, drawing attention to the wide 
parameter space that dwarf starbursts cover. 

As an example of the complex nature of radio continuum emission in these
low-mass galaxies, consider NGC\,625, NGC\,1569, and NGC\,1705. All three 
systems have large starburst regions with high \halpha\ equivalent widths, and
all have outflows detected in x-ray, uv, optical or radio observations [see 
{Heckman \etal\ (2001)}\nocite{heckman01a} for NGC\,1705; {Martin, Kobulnicky 
\& Heckman (2002)}\nocite{martin02} for NGC\,1569; {Cannon \etal\ 
(2004a)}\nocite{cannon04a} for NGC\,625].  NGC\,625 and NGC\,1569 show thermal 
radio emission while NGC\,1705 shows no thermal emission to reasonable 
detection levels.  Although outflow is suggested to explain the rapid cessation
of star formation in NGC\,1705 \citep{johnson03a}, strong outflows exist in 
both NGC\,625 and NGC\,1569, and current star formation does not appear to be 
significantly affected (although the star formation rates in both systems have 
declined over the last $\sim$ 25 Myr).  If outflow and its effects on future 
star formation are responsible for the variation in radio continuum properties 
of these star-forming galaxies, then its nature is a complicated one that 
depends on characteristics of the ISM that have not yet been included in models
of the formation and evolution of outflows from these systems.

\subsection{Radio Continuum Emission in Wolf-Rayet Galaxies}
\label{S4.2}

Each of the above systems displays broad W-R emission features in its optical 
spectrum ({Schaerer, Contini \& Pindao 1999b}\nocite{schaerer99b}; {Skillman
\etal\ 2003b}\nocite{skillman03b}) except for NGC\,1705 (Lee \& Skillman 2004,
in preparation).  Based on the lifetimes discussed in \citet{conti91} and 
{Schaerer \etal\ (1999a)}\nocite{schaerer99a}, these features imply recent ($<$
6 Myr) star formation bursts with short durations ($\sim$ 2 -- 4 Myr).  Of 
course these spectral features arise from individual star formation regions and
indicate the ages of that particular stellar population.  However, they are not
necessarily indicative of the entire stellar population in the galaxies, as 
discussed in \citet{cannon03}.  Active star formation can pervade the disks of 
low-mass galaxies for extended periods of time, and the W-R features simply 
highlight active regions at the current epoch.  It is important to keep this in
mind for galaxies at distances where a resolved stellar population analysis is 
not available. 

Since the bulk of W-R galaxies are more distant than the sample discussed 
above, the number of studies comparing their resolved stellar populations with 
high-resolution radio continuum maps is small.  The largest sample to date was 
compiled by {Beck, Turner, \& Kovo (2000)}\nocite{beck00}, where nine 
relatively nearby W-R galaxies were observed at multiple frequencies.  The 
results of that study suggest that thermal radio continuum emission is common 
from W-R galaxies, and in many cases, so is the appearance of positive-index 
embedded sources.  This can be contrasted with the \HII\ galaxy sample of 
\citet{deeg93}, where most galaxies show nonthermal spectral indices.  
\citet{beck00} interpret the prominent free-free emission in W-R galaxies to 
the youth of these systems, while the strong synchrotron components found in 
the sample of \citet{deeg93} suggest that the star formation events in these 
systems are somewhat older.  This comparison could be problematic, however, 
given the differences in observing frequency and synthesized beam between the 
two investigations. 

This limited sample of W-R galaxies appears to support the interpretation of 
the spectral index as a rough age indicator.  In this paradigm, NGC\,1705 not 
appearing as a W-R galaxy in the optical places it more correctly in the class 
of \HII\ galaxies, and its very low radio continuum luminosity is not as 
striking as when compared to more active W-R and dwarf starburst galaxies. This
scenario is certainly not robust, however, and the effects of feedback in 
regulating the evolution of radio continuum emission in dwarf starburst 
galaxies needs more detailed attention.  With the imminent completion of the 
Expanded Very Large Array\footnote{See http://www.aoc.nrao.edu/evla/}, 
providing a factor of $>$ 10 improvement in spatial resolution and sensitivity,
the sample of W-R galaxies can be greatly expanded, and the nature of radio 
continuum emission therein can be explored in much more detail.

\section{Conclusions}
\label{S5}

We have presented new VLA radio continuum imaging, in the BnA and CnB arrays at
L, C and X-bands, of the nearby dwarf starburst galaxy NGC\,625. The global 
spectral index is nearly flat, suggesting that thermal emission dominates the 
radio continuum luminosity.  Examination of the spectral indices of individual 
star formation regions shows a mix of thermal and nonthermal processes.  The 
highest-surface brightness \HII\ regions NGC\,625\,A and NGC\,625\,B show 
strong free-free emission, while the less-luminous \HII\ region NGC\,625\,C is 
dominated by nonthermal synchrotron radiation. At $\sim$ 2\arcsec\ ($\sim$ 40 
pc) resolution, we do not find any deeply-embedded sources in NGC\,625. 
However, we do find (with only 3\,$\sigma$ significance) one low-luminosity 
source that has no obvious optical counterpart and that is located in the 
region of highest optical extinction. 

We interpret the mix of thermal and nonthermal emission from the main star 
formation regions as an age progression, with free-free emission arising from 
\HII\ regions with ages $<$ 10 Myr, and synchrotron radiation prominent in 
regions older than this.  Comparing to a limited sample of well-studied dwarf 
starburst and W-R galaxies, we find that global thermal emission is common in 
these types of systems.  The spectral indices are typically flatter than in 
\HII\ galaxies, and this is again suggestive that the dominant free-free 
component in dwarf starburst and W-R galaxies implies younger starburst ages 
than in more evolved systems where synchrotron emission dominates.  

The importance of small-scale processes (i.e., feedback) in regulating the 
evolution of emission in these systems is highlighted.  While a simple age 
interpretation is suitable for most of this small nearby sample, more complex 
interpretations are needed when comparing multiwavelength properties of various
galaxies.  The formation and evolution of outflows is an important parameter, 
but is relatively poorly constrained by observations.  More detailed modeling 
and higher-resolution observations of larger samples of dwarf starburst and W-R
galaxies would be beneficial for understanding the evolution of the radio 
continuum properties of these important star-forming systems.

\acknowledgements

The authors appreciate the careful reading and helpful comments of an anonymous 
referee that improved this manuscript.  J.\,M.\,C. is supported by NASA 
Graduate Student Researchers Program (GSRP) Fellowship NGT 5-50346.  E.\,D.\,S.
is grateful for partial support from NASA LTSARP grant NAG5-9221 and the 
University of Minnesota. This research has made use of the NASA/IPAC 
Extragalactic Database (NED) which is operated by the Jet Propulsion 
Laboratory, California Institute of Technology, under contract with the 
National Aeronautics and Space Administration, and NASA's Astrophysics Data 
System. 

\clearpage


\clearpage

\clearpage

\clearpage

\clearpage

\clearpage

\clearpage
\begin{deluxetable}{ccc}
\tabletypesize{\scriptsize}
\tablecaption{Basic Parameters of NGC\,625}
\tablewidth{0pt}
\tablehead{
\colhead{Property}         
&\colhead{Value} 
&\colhead{Reference}}
\startdata
Distance (Mpc)  	&3.89\,$\pm$\,0.22		&\citet{cannon03}\\
M$_B$			&$-$16.3		&\citet{marlowe97}\\
Gal. Lat. (\degree)	&$-$73.1		&$--$\\
Foreground E(B$-$V) &0.016		
&{Schlegel, Finkbeiner, \& Davis (1998)}\nocite{schlegel98}\\
12\,$+$\,log(O/H) 	&8.14\,$\pm$\,0.02		&\citet{skillman03b}\\
Current SFR (\msun\,yr$^{-1}$)  &0.05  		&\citet{skillman03a}\\
\HI\ Mass (10$^8$ \msun) &1.1\,$\pm$\,0.1		&{Cannon \etal\ 
(2004a)}\nocite{cannon04a}\\
V$_{Helio}$ (\kms)	&413\,$\pm$\,5 		&{Cannon \etal\ 
(2004a)}\nocite{cannon04a}\\
V$_{20}$ (\kms)		&95\,$\pm$\,2		&{Cannon \etal\ 
(2004a)}\nocite{cannon04a}\\
V$_{50}$ (\kms)		&62\,$\pm$\,2		&{Cannon \etal\ 
(2004a)}\nocite{cannon04a}\\
\enddata
\label{t1}
\end{deluxetable}

\clearpage
\begin{deluxetable}{ccccccc}
\tabletypesize{\scriptsize}
\tablecaption{VLA Observations of NGC\,625 for Program AC\,702}
\tablewidth{0pt}
\tablehead{
\colhead{Array}         
&\colhead{Observation} &\colhead{Frequency} 
&\colhead{T$_{INT}$} &Flux &Phase &Phase Calibrator\\
\colhead{Configuration} 
&\colhead{Date(s)}
&\colhead{(GHz)}   
&\colhead{(Minutes)} &Calibrator &Calibrator &F$_{\nu}$ (Jy) }
\startdata
BnA &2003 Oct 9  &1.40 & 178 &3c\,147 &0155$-$408  &2.1\,$\pm$\,0.10\\
BnA &2003 Oct 10 &4.86 & 178 &3c\,147 &0155$-$408 &1.09\,$\pm$\,0.06\\
BnA &2003 Oct 11 &8.46 & 178 &3c\,147 &0155$-$408 &0.78\,$\pm$\,0.04\\
CnB &2004 Feb 7 &4.86  & 178 &3c\,147 &0155$-$408 &1.08\,$\pm$\,0.06\\
CnB &2004 Feb 8 &8.46  & 178 &3c\,147 &0155$-$408 &0.78\,$\pm$\,0.04\\
\enddata
\label{t2}
\end{deluxetable}

\clearpage
\begin{deluxetable}{lccccc}
\tabletypesize{\scriptsize}
\tablecaption{Properties of Radio Continuum Emission}
\tablewidth{0pt}
\tablehead{
\colhead{Parameter}         
&\colhead{NGC\,625\,A\tablenotemark{a}}
&\colhead{NGC\,625\,B\tablenotemark{a}} 
&\colhead{NGC\,625\,C\tablenotemark{a}}
&\colhead{NGC\,625\,D}
&\colhead{NGC\,625 Total Galaxy}}
\startdata
\vspace{0.1 cm}
R.A. (J2000)\tablenotemark{b} &01:35:07.0 &01:35:06.0 &01:35:05.0 
&01:35:04.9 &01:35:04.6\\ 
Dec. (J2000)\tablenotemark{b} &$-$41:26:12.0 &$-$41:26:11.0 &$-$41:26:09.5
&$-$41:26:17.2 &$-$41:26:10.0\\ 
S$_{\rm L\,Band}$ (mJy) &7.4\,$\pm$\,0.8 &0.80\,$\pm$\,0.08 &0.28\,$\pm$\,0.03
&0.09\,$\pm$\,0.02 &9.9\,$\pm$\,1.0\tablenotemark{c}\\
S$_{\rm C\,Band}$ (mJy) &6.7\,$\pm$\,0.7 &0.77\,$\pm$\,0.08 &0.20\,$\pm$\,0.02
&0.08\,$\pm$\,0.02 &8.9\,$\pm$\,0.9\tablenotemark{c}\\
S$_{\rm X\,Band}$ (mJy) &6.3\,$\pm$\,0.7 &0.60\,$\pm$\,0.06 &0.15\,$\pm$\,0.02
&0.06\,$\pm$\,0.02 &7.6\,$\pm$\,0.8\tablenotemark{c}\\
$\alpha$\tablenotemark{d} &$-$0.09\,$\pm$\,0.08 &$-$0.14\,$\pm$\,0.08 
&$-$0.33\,$\pm$\,0.08 &$-$0.22\,$\pm$\,0.19 &$-$0.13\,$\pm$\,0.08\\ 
T$_e$ (K)\tablenotemark{e} &10900$^{+115}_{-109}$ &10460$^{+213}_{-193}$ 
&12810$^{+460}_{-395}$ &--- &---\\
Q$_{Lyc}$ (10$^{50}$ photons\,sec$^{-1}$)\tablenotemark{f} &85 &9.2 
&--- &---  &---\\ 
SFR (\msun\,yr$^{-1}$)\tablenotemark{f,g} &0.026 &0.003 &--- &--- &---\\
\halpha\ Flux (10$^{-14}$ erg\,sec$^{-1}$\,cm$^{-2}$)\tablenotemark{h}
&211\,$\pm$\,10 &112\,$\pm$\,6 &7.8\,$\pm$\,0.4 &$<$\,0.9 &350\,$\pm$\,18\\
Implied A$_{H\alpha}$ (mag.)\tablenotemark{f} &1.1\,$\pm$\,0.3 &0.0\,$\pm$\,0.2
&--- &$>$ 2.0\tablenotemark{i} &---\\
\enddata
\label{t3}
\tablenotetext{a}{Adopting the nomenclature of 
\citet{cannon03}; see Figures~\ref{figcap1}(b) and \ref{figcap5}(b).}
\tablenotetext{b}{Central position of radio continuum emission peaks are 
derived from Gaussian fitting.  The coordinates of the NGC\,625 system are 
taken from the NASA Extragalactic Database.} 
\tablenotetext{c}{Measured using an identical aperture around the entire system
for each frequency, and integrating all emission above the 3\,$\sigma$ level.}
\tablenotetext{d}{$\alpha$ in the relation \snu\ $\sim\ \nu^{\alpha}$; this 
is derived by a least-squares power law fit to each of the data points 
measured.}
\tablenotetext{e}{Electron temperatures are adopted from \citet{skillman03b}.}
\tablenotetext{f}{Only applicable to sources with a thermal emission origin; 
see \citet{condon92}.  Q$_{Lyc}$ and SFR are the averages of each measure from 
the three available frequencies.}
\tablenotetext{g}{Star formation rate for stars more massive than 5\,\msun; see
\citet{condon92}.}
\tablenotetext{h}{Derived in \citet{cannon03}.}
\tablenotetext{i}{Extinction toward source NGC\,625\,D is also calculated, 
since the error on the spectral index may allow a thermal slope.  If the 
\halpha\ flux has been underestimated by $\sim$ 50\%, then this implied 
extinction falls to \gsim\ 1.1 magnitudes (see discussion in \S~\ref{S3.4}).}
\end{deluxetable}

\clearpage
\begin{deluxetable}{cccccc}
\tabletypesize{\scriptsize}
\tablecaption{Property Comparison of Selected Nearby Dwarf Starburst Galaxies}
\tablewidth{0pt}
\tablehead{
\colhead{Property\tablenotemark{a}}         
&\colhead{NGC\,625} 
&\colhead{NGC\,1569}
&\colhead{NGC\,1705}
&\colhead{He\,2-10}
&\colhead{NGC\,5253}}
\startdata
Distance (Mpc) &3.89\,$\pm$\,0.22 (1) &2.2\,$\pm$\,0.6 (2) &5.1\,$\pm$\,0.6 (3)
&9.0 (4)\tablenotemark{b} &4.1\,$\pm$\,0.2 (5)\\

\HI\ Mass (\msun) &1.1$\times$10$^8$ (6) &1.3$\times$10$^8$ (7) 
&1.5$\times$10$^8$ (8) &3.3$\times$10$^8$ (9) &1.4$\times$10$^8$ (10)\\

M$_B$ &$-$16.3 &$-$17.0 &$-$16.3 &$-$17.4 &$-$17.1\\

E(B-V)\tablenotemark{c} &0.016 &0.695 &0.008 &0.112 &0.074\\

12+log(O/H) &8.14\,$\pm$\,0.02 (11) &8.19$\pm$0.02 (12) &8.22\,$\pm$\,0.06 (13)
&8.93 (14)\tablenotemark{b} &8.15$\pm$0.04 (15)\\

Current SFR (\msun\ yr$^{-1}$)\tablenotemark{d} &0.05 (1) &0.1 (16) &0.1 (17) 
&1.4 (18) &0.2 (16)\\

Global spectral index, $\alpha$ &$-$0.13\,$\pm$\,0.08 (19) 
&$-$0.25 (20)\tablenotemark{b} &--- &$-$0.54\,$\pm$\,0.03 (4) 
&0.0\,$\pm$\,0.1 (21)\\

X-Ray Luminosity\\
(10$^{36}$ erg/sec) &100 (22) &800 (23) &120 (24) &1200 (25)\tablenotemark{e} 
&650 (26)\\
\enddata
\tablerefs{
1 - \citet{cannon03};
2 - \citet{israel88a};
3 - \citet{tosi01};
4 - \citet{kobulnicky99c};
5 - \citet{saha95};
6 - {Cannon \etal\ (2004a)}\nocite{cannon04a};
7 - \citet{stil02b};
8 - \cite*{meurer98};
9 - \citet{sauvage97};
10 - \citet{reif82};
11 - \citet{skillman03b};
12 - \citet{kobulnicky97b};
13 - Lee \& Skillman, in preparation;
14 - \citet{kobulnicky99b};
15 - \citet{kobulnicky97a};
16 - \citet{martin98};
17 - \citet{marlowe97};
18 - \citet{mendez99};
19 - This work; 
20 - \citet{israel88b};
21 - \citet{turner98};
22 - \citet{bomans98};
23 - \citet{martin02};
24 - \citet{hensler98};
25 - Martin \& Kobulnicky, in preparation;
26 - \citet{martin95}}
\tablenotetext{a}{The number in parentheses after some entries corresponds to 
the reference from which that data point was taken.}
\tablenotetext{b}{No uncertainty quoted.}
\tablenotetext{c}{Foreground extinction only, drawn from \citet{schlegel98}.}
\tablenotetext{d}{Applying \halpha\ fluxes from the works cited, and the 
conversion to star formation rate from \citet*{kennicutt94}.}
\tablenotetext{e}{Diffuse gas component (0.3 - 6.0 keV) only.}
\label{t4}
\end{deluxetable}
\end{document}